%% file: axaf.tex
\def\puncspace{\ifmmode\,\else{\ifcat.\C{\if.\C\else\if,\C\else\if?\C\else%
\if:\C\else\if;\C\else\if-\C\else\if)\C\else\if/\C\else\if]\C\else\if'\C%
\else\space\fi\fi\fi\fi\fi\fi\fi\fi\fi\fi}%
\else\if\empty\C\else\if\space\C\else\space\fi\fi\fi}\fi}
\def\SP{\let\\=\empty\futurelet\C\puncspace}

\def\h1{$h^{-1}$\SP}

\def\etal{{\it et al.\/}\ }

\def\eg{{\it e.g.\/}\rm,\ }
\def\lsim{~\rlap{$<$}{\lower 1.0ex\hbox{$\sim$}}}
\def\gsim{~\rlap{$>$}{\lower 1.0ex\hbox{$\sim$}}}
\def\void#1{{}}
\def\u{$U$}
\def\b{$B$}
\def\v{$V$}
\def\r{$R$}
\def\i{$I$}
\def\j{$J$}
\def\h{$H$}
\def\k{$Ks$}
\documentclass{aa}
\usepackage{graphicx}

\usepackage{times}
\usepackage{mathptm}
\fontfamily{ptmcm}\selectfont
\usepackage{graphicx}

\begin{document}

   \thesaurus{02 (04.19.1; 04.03.1; 08.07.1; 11.07.1; 12.03.3)}
   \title{ESO Imaging Survey}

   \subtitle{AXAF Field: Deep Optical-Infrared Observations, Data
Reduction and Photometry}

\author{  
 R. Rengelink \inst{1}
 \and M. Nonino \inst{1,2}
 \and  L. da Costa\inst{1}
 \and S. Zaggia \inst{1,3,4} 
 \and T. Erben\inst{1,5}
 \and C. Benoist\inst{1,5} 
 \and A. Wicenec\inst{1}
 \and M. Scodeggio \inst{1}
 \and L. F. Olsen\inst{1,6}
 \and D. Guarnieri\inst{7} 
 \and E. Deul\inst{1,8}
 \and R. Hook\inst{9} 
 \and A. Moorwood \inst{1}
 \and R. Slijkhuis\inst{1}  
 }

\institute{
European Southern Observatory, Karl-Schwarzschild-Str. 2,
D--85748 Garching b. M\"unchen, Germany 
 \and Osservatorio Astronomico di Trieste, Via G.B. Tiepolo 11, I-31144
Trieste, Italy
\and Dipartimento di Astronomia, Univ. di Padova, vicolo
dell'Osservatorio 5, I-35125, Padova, Italy
\and Osservatorio Astronomico di Capodimonte, via Moiariello 15,
 I-80131, Napoli, Italy 
 \and Max-Planck Institut f\"ur Astrophysik, Postfach 1523 D-85748, 
 Garching bei M\"unchen, Germany
 \and Astronomisk Observatorium, Juliane Maries Vej 30, DK-2100 Copenhagen, 
  Denmark
 \and Osservatorio Astronomico di Pino Torinese, Strada Osservatorio
 20, I-10025 Torino, Italy
  \and Leiden Observatory, P.O. Box 9513, 2300 RA Leiden, The
  Netherlands 
  \and Space Telescope -- European Coordinating Facility,
 Karl-Schwarzschild-Str. 2, D--85748 Garching b. M\"unchen, Germany }

   \date{Received ; accepted}

   \maketitle

   \begin{abstract}

This paper presents ground-based data obtained from deep optical and
infrared observations carried out at the ESO 3.5 New Technology
Telescope (NTT) of a field selected for its low HI content for deep
observations with AXAF.  These data were taken as part of the ESO
Imaging Survey (EIS) program, a public survey conducted in preparation
for the first year of operation of the VLT. Deep CCD images are
available for five optical passbands, reaching $2\sigma$ limiting
magnitudes of $U_{AB}\sim$ 27.0, $B_{AB}\sim$ 27, $V_{AB}\sim$ 26.5,
$R_{AB}\sim$ 26.5 and $I_{AB}\sim$ 26. An area of $\sim 56$ square
arcmin is covered by $UBVR$ observations, and $\sim$ 30 square arcmin
also in \i.  The infrared observations cover a total area of $\sim$ 83
square arcmin, reaching $J_{AB}\sim 24.5$ and $K_{AB}\sim$ 23.5. This
paper describes the observations and data reduction. It also presents
images of the surveyed region and lists the optical and infrared
photometric parameters of the objects detected on the co-added images
of each passband, as well as multicolor optical and infrared catalogs.
These catalogs together with the astrometrically and photometrically
calibrated co-added images are public and can be retrieved from the
URL ``http://www.eso.org/eis''. This data set completes the ESO
Imaging Survey program sixteen months after it began in July 1997.

      \keywords{catalogs -- surveys -- stars:general --
                galaxies:general -- cosmology:observations 
               }
   \end{abstract}

%

\section{Introduction}

One of the main goals of the ESO Imaging Survey (EIS, Renzini \& da
Costa 1997) has been to carry out deep, multicolor observations in the
optical and infrared passbands over a relatively large area ($\sim$
200 square arcmin) to produce faint galaxy samples (EIS-DEEP). The
primary objective is to use the color information to estimate
photometric redshifts, and identify galaxies likely to be in the
poorly sampled $1 \lsim z<2$ redshift interval or Lyman-break
candidates at $z\gsim2.5$, all of which are interesting targets for
follow-up spectroscopic observations with the VLT.

The first part of EIS-DEEP consisted of optical and infrared
observations of the Hubble Deep Field South (HDF-S). The results  have
recently been reported and the data made public world-wide (da Costa
\etal 1998, hereafter paper~I). However, as pointed out in that paper, 
EIS-DEEP observations were also carried out over a region (hereafter
the AXAF field) around $\alpha=03^h32^m28^s $ and
$\delta=-27^\circ48'30''$ selected for its low HI column density and
where deep X-ray observations will be conducted with AXAF (Giacconi
1998). This field, in contrast to HDF-S, is remarkably devoid of
bright stars and therefore particularly suitable for deep ground-based
imaging. The aim was to cover about 100 square arcmin in five optical
($UBVRI$) and two infrared passbands ($JKs$) to similar depth of the
HDF-S observations, thereby providing another field from which targets
can be selected and observed in a different period of the year.

In the present paper the data from the AXAF field observations are
reported.  In section~\ref{obs}, the observations and data reduction
are discussed, and co-added images for the different fields and
passbands are presented. In section~\ref{cats}, the basic photometric
parameters of sources detected in each passband are listed as well as
optical, infrared and optical-infrared multicolor catalogs. Using
these catalogs the characteristics of the data are explored in
section~\ref{results}.  Finally, a preliminary list of high-z galaxy
candidates is selected.  A brief summary is presented in
section~\ref{sum}.

\section {Observations and Data Reduction}
\label{obs}

The original goal of the EIS-DEEP observations of the AXAF field was
to observe four adjacent and contiguous regions centered on the
nominal position selected for the X-ray observations, where the
sensitivity and image quality of the X-ray images are expected to be
the best.  Therefore, the pointings of the optical and infrared
cameras were chosen in such a way as to have the X-ray center in a
region where the sensitivity of the mosaic, built from different
pointings from the ground-based observations, is highest, taking into
account both the physical characteristics of the optical and infrared
cameras used and the overlap between the images. The pointings adopted
in the construction of the optical and infrared mosaics are listed in
Table~\ref{tab:pointings}. As discussed below, while the infrared
mosaic is essentially complete in $J$ and $Ks$, the optical is only
half-way complete due to time and weather constraints.

\begin{table}[t]
\caption{SUSI2 and SOFI Pointings (J2000.0)}
\begin {tabular}{lcccc}
\hline \hline
Field & $\alpha_{susi2}$ & $\delta_{susi2}$ & $\alpha_{sofi}$ &
$\delta_{sofi}$ \\
\hline
AXAF1 &  03:32:16.7 & -27:46:00.0 & 03:32:17.5 &-27:46:10.0 \\
AXAF2 &  03:32:38.0 & -27:46:00.0 & 03:32:37.5 &-27:46:10.0 \\
AXAF3 &   -         &   -          & 03:32:17.5 & -27:50:35.0 \\
AXAF4 &   -         &   -          & 03:32:37.5 & -27:50:35.0 \\
\hline
\hline
\label{tab:pointings}
\end{tabular}
\end{table}

The optical observations were carried out using the SUSI2 camera
(D'Odorico \etal 1998) at the f/11 Nasmyth focus A of the New
Technology Telescope (NTT). The camera consists of two thinned,
anti-reflection coated, $2k\times4k$, $15\mu m$ pixel EEV CCDs (ESO
\#45, and \#46), with the long side aligned in the north-south
direction, leading to a field of view of $5.46\times5.46$ square
arcmin.  The observations were carried out using a $2\times2$ binning,
yielding a scale of $0.16$ arcsec per pixel. The gap separating the
two CCDs corresponds to $\sim 8$ arcsec on the sky.  To minimize the
effect of the inter-chip gap. the observations were carried out using
the same dithering pattern as described in paper~I.

The optical observations were carried out in the period
August-November 1998, using broad-band $UBVRI$ filters (ESO \# 810,
811, 812, 825, and 814, see SUSI2 web page). With exception of the \r\
filter, these are the same filters used in the HDF-S observations. A
total of 12 nights were allocated for optical EIS-DEEP observations
which were split between the HDF-S and AXAF fields. However, because
of poor weather conditions it was not possible to complete the program
as originally envisioned.  Complete observations in all optical
passbands are available for AXAF1, while for AXAF2 \i-band data are
missing.

Table~\ref{tab:exposures} summarizes the observations, listing for
each field and passband the total integration time, the number of
exposures, the range of seeing as measured on individual exposures,
the full-width at half-maximum (FWHM) of the point spread function
(PSF) in the final co-added image, and the estimated $1\sigma$
limiting isophote within a 1 square arcsec area.  Single exposures
ranged from 800 sec ($U$) to 250 sec ($R$).

\begin{table}[t]
\caption{Summary of Optical Observations}
\begin{tabular}
{lrcccc}
\hline\hline
Filter & $t_{total}$ &  $N_f$ & seeing       & FWHM$^1$  & $\mu_{lim}^1$\\
       &             &        &        range &       &            \\
       &    (sec)    &        &  (arcsec)    & (arcsec) &  (mag arcsec$^{-2}$) \\
\hline
AXAF1       &      &       &           &       \\
\hline
   \u & 17000 &   21  & 0.8-1.0 & 0.90    & 27.81 \\
   \b &  6600 &   22  & 0.7-1.0 & 1.10    & 28.76 \\
   \v &  5500 &   22  & 1.0-1.2 & 0.88    & 28.46 \\
  \r &  5500  &   22  & 0.7-1.0 & 0.89    & 28.06 \\
  \i &  12600 &   21  & 0.8-1.4 & 1.31    & 27.51 \\
\hline
AXAF2       &      &       &           &       & \\
\hline
   \u & 13000 &   16  & 0.7-1.2 & 0.90    & 27.81 \\
   \b &  5400 &   18  & 1.0-1.2 & 1.10    & 28.76 \\
   \v &  5500 &   22  & 0.7-1.1 & 0.88    & 28.46 \\
   \r &  5500 &   22  & 0.7-1.1 & 0.89    & 28.06 \\
\hline\hline
\end{tabular}

$^1$ Values determined from AXAF1+2 mosaic.

\label{tab:exposures}
\end{table}

As described in paper~I, analysis of the Landolt standards yields the
following estimates for the accuracy of the zero-points: $\pm$ 0.1~mag
in \u; $\pm$0.03~mag in \b; $\pm$0.03~mag in \v; $\pm$0.04~mag in \r;
and $\pm$0.05~mag in \i. These results apply to both chips, with the
relative zero-point difference between the chips being smaller than
their estimated errors. Only for the \u\ and \b\ filters there are
significant color terms relative to the Johnson system.  The
photometric accuracy of the EIS-DEEP data was evaluated in paper~I by
comparing them with those available from other independent
observations of the HDF-S field.  As described in paper~I the
agreement is excellent with the scatter being fully accounted for by
the measurement errors estimated from SExtractor.

\begin{figure}[t]
\resizebox{0.40\textwidth}{!}{\includegraphics{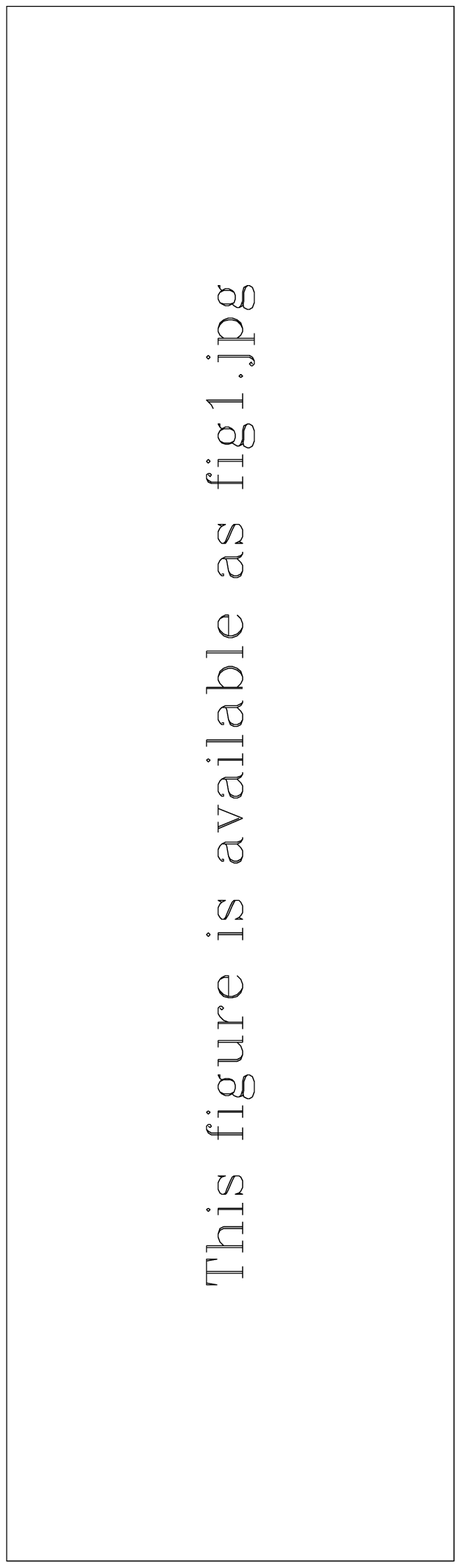}}
\caption{Final  co-added mosaic  ($11.2 \times 5.6$ square arcmin) for
 each of the optical passbands showing from top to bottom the \u,
\b, \v, \r\ and \i. Note that only one field is avaliable in \i.}
\label{fig:opmosaic}
\end{figure}

The measured magnitudes were corrected for galactic absorption, using
$E(B-V)=0.009$ as derived from Schlegel, Finkbeiner \& Davis (1998),
yielding $A_U=0.05$~mag, $A_B$=0.03~mag, $A_V$=0.02~mag,
$A_R$=0.02~mag and $A_I$=0.01~mag.  To facilitate the comparison with
other authors, all magnitudes given below have been converted to the
$AB$ system using the following relations: $U_{AB} = U + 0.82; B_{AB}=
B - 0.06; V_{AB}= V; R_{AB}= R + 0.17;$ and $I_{AB}=I + 0.42$, unless
specified otherwise.

A total of 180 science frames were obtained for the AXAF field and
reduced using standard IRAF tasks. A complete description of the
various reduction steps can be found in paper~I. After the frames were
corrected for all instrumental effects, an eye-inspection was carried
out to mask out features such as satellite tracks.

\begin{figure}[t]
\resizebox{0.45\textwidth}{!}{\includegraphics{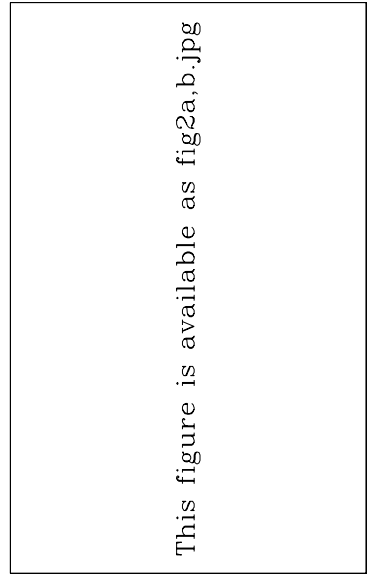}}
\caption{Final co-added infrared mosaic \j\ (upper
panel) and \k\ (lower panel). The images are about $9.0 \times
9.0$ square arcmin. }
\label{fig:irmosaic}
\end{figure}

Infrared observations were obtained using the SOFI camera (Moorwood,
Cuby \& Lidman 1998) also at the NTT. SOFI is equipped with a Rockwell
1024$^2$ detector that, when used together with its large field
objective, provides images with a pixel scale of 0.29 arcsec, and a
field of view of about $4.9 \times 4.9$ square arcmin. The infrared
pointings, listed in Table~\ref{tab:pointings} were chosen so as to
take into account the smaller field of view of SOFI and produce a
contiguous mosaic slightly smaller than that in the optical.  The
final infrared mosaic has an area of $83$ square arcmin. The jittering
pattern used was the same as described in paper~I. Individual
observations comprised sixty one-minute exposures, with each exposure
being the average of six ten-second sub-exposures.

\begin{table}
\caption{Summary of Infrared  Observations}
\begin{tabular}{lrcccc}
\hline\hline
Filter & $t_{total}$   & $N_f$ &seeing  & FWHM$^1$ &   $ \mu_{lim}^1$\\
       &               &       & range  &      &               \\
       &    (sec)    &        &  (arcsec)    & (arcsec) & (mag arcsec$^{-2}$)\\
\hline
AXAF1  &      &             &      &   & \\
\hline
  \j & 10800 & 180 & 0.8-1.2 & 0.99 & 25.32  \\
  \k & 10800 & 180 & 0.8-1.1 & 0.95 & 23.83 \\
\hline
AXAF2  &      &            &       &   & \\
\hline
   \j & 10800 & 180 & 0.7-0.9 & 0.99 & 25.32 \\
   \k & 10800&  180 & 0.8-1.0 & 0.95 & 23.83 \\
\hline
AXAF3  &      &            &       &   & \\
\hline
   \j & 10800 & 180 & 0.7-0.9 & 0.99 & 25.32 \\
   \k & 10800&  180 & 0.8-1.3 & 0.95 & 23.83 \\
\hline
AXAF4  &      &            &       &   & \\
\hline
   \j & 10800 & 180 & 0.7-0.8 & 0.99 & 25.32 \\
   \k & 7200  & 120 & 0.7-1.0 & 0.95 & 23.83 \\
\hline\hline
\end{tabular}

$^1$ Values determined from combined AXAF1-4 mosaic.

\label{tab:obsir}
\end{table}

Infrared observations in the $J$ and $Ks$ bands were obtained during
the same period as the optical data. Total integration times, the
number of frames, the seeing range, the FWHM measured on the co-added
images, and the estimated $1\sigma$ limiting isophote within one
square arcsec are given in Table~\ref{tab:obsir}.

During all nights infrared standards taken from Persson (1997) were
observed. From the photometric solutions the errors in the absolute
photometric zero-points are found to be: $\pm$ 0.05~mag in \j\ and
$\pm$ 0.05~mag in \k. These magnitudes were also converted to the $AB$
system, using: $J_{AB}= J + 0.89$ and $K_{AB}= Ks + 1.86$.

A total of 1380 SOFI science frames in the AXAF field were reduced, as
detailed in paper~I, utilizing the program {\em jitter} from the
astronomical data-reduction package {\em eclipse}, written by
N. Devillard (Devillard, 1998).

After removing all the instrumental signatures, both optical and
infrared images were input to the EIS pipeline for astrometric
calibration using the USNO-A V1.0 catalog as reference. In the case of
SUSI2, independent astrometric solutions were found for the two chips.
Before the astrometric calibration, images were processed by
SExtractor using a high detection threshold to measure the size of the
PSF of each frame, to create weight maps for co-addition and to flag
cosmic rays and defects.  Images in the same passband were then
co-added using the "drizzle" method, originally created to handle HST
images (Fruchter \& Hook 1998), at the same resolution as the original
images. In the process of co-addition, images taken in photometric
nights were used as reference and all other images were corrected to
have the same instrumental magnitudes after extinction
correction. Finally, the absolute zero-point was determined and stored
in the header of the co-added image. In addition, in order to allow
the production of optical-infrared multi-color catalogs based on a
single detection image ($\chi^2$-image), a co-added image of the
optical data matching the infrared image resolution was also created
over the area they overlap.

\begin{figure}[t]
\resizebox{0.45\textwidth}{!}{\includegraphics{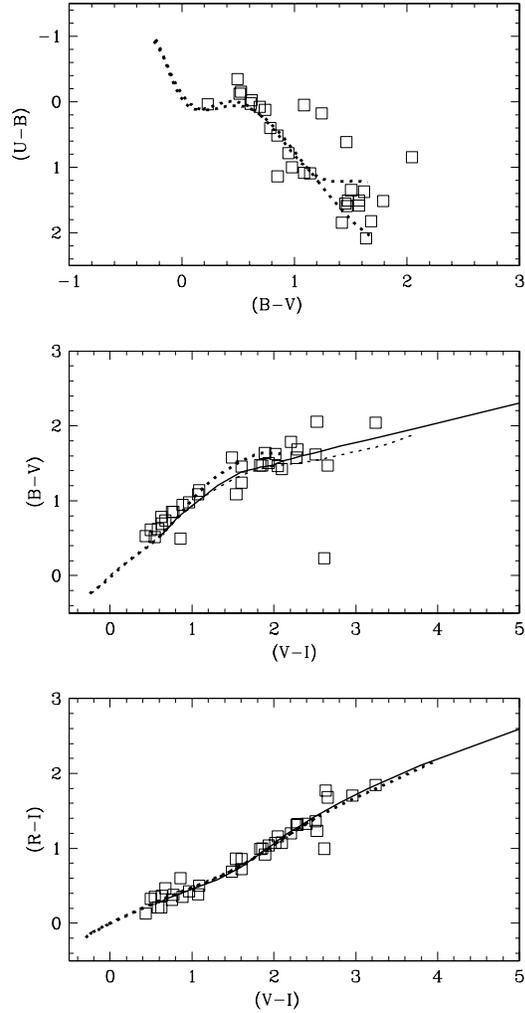}}
\caption{Color-color diagrams for point sources in the AXAF field
compared with empirical determinations (dashed line) and model
predictions (solid line). Only $5\sigma$ detections with a stellarity
index $> 0.95$ are included.}
\label{fig:starcol}
\end{figure}

\begin{figure*}
\resizebox{0.6\textwidth}{!}{\includegraphics{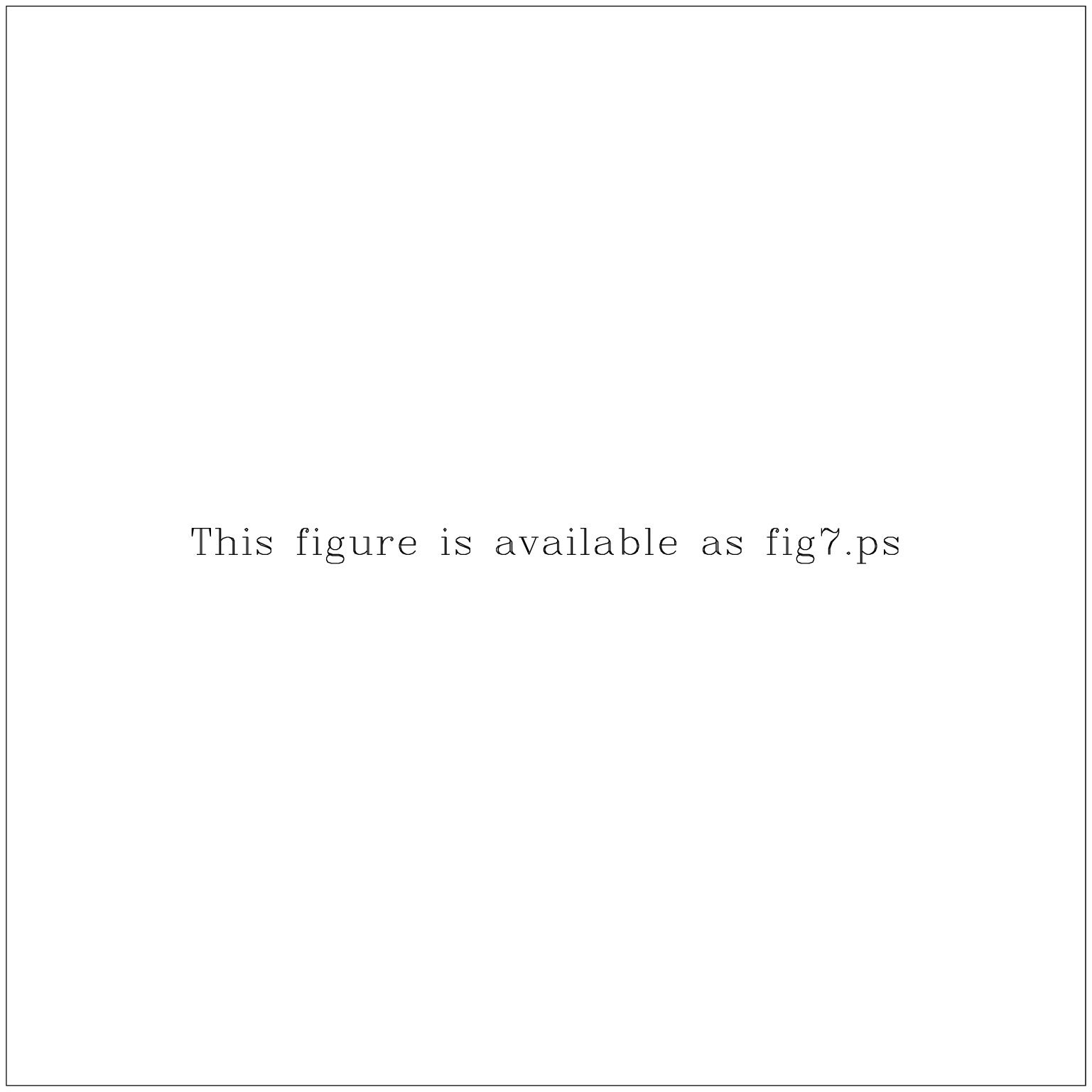}}
\caption{Plate 1. True-color image of the AXAF optical mosaic  based on 
four passbands. The blue channel is represented by the
\u+\b\ image, the green channel by the
\v\ image, and the red channel by the
\r\ image. The edges of the field have been trimmed to exclude the
imprint of the dithering pattern. This color
image covers an area of approximately $10.5\times5.3$ square arcmin.}
\label{fig:plate1}
\end{figure*}

\begin{figure*} 
\resizebox{0.9\textwidth}{!}{\includegraphics{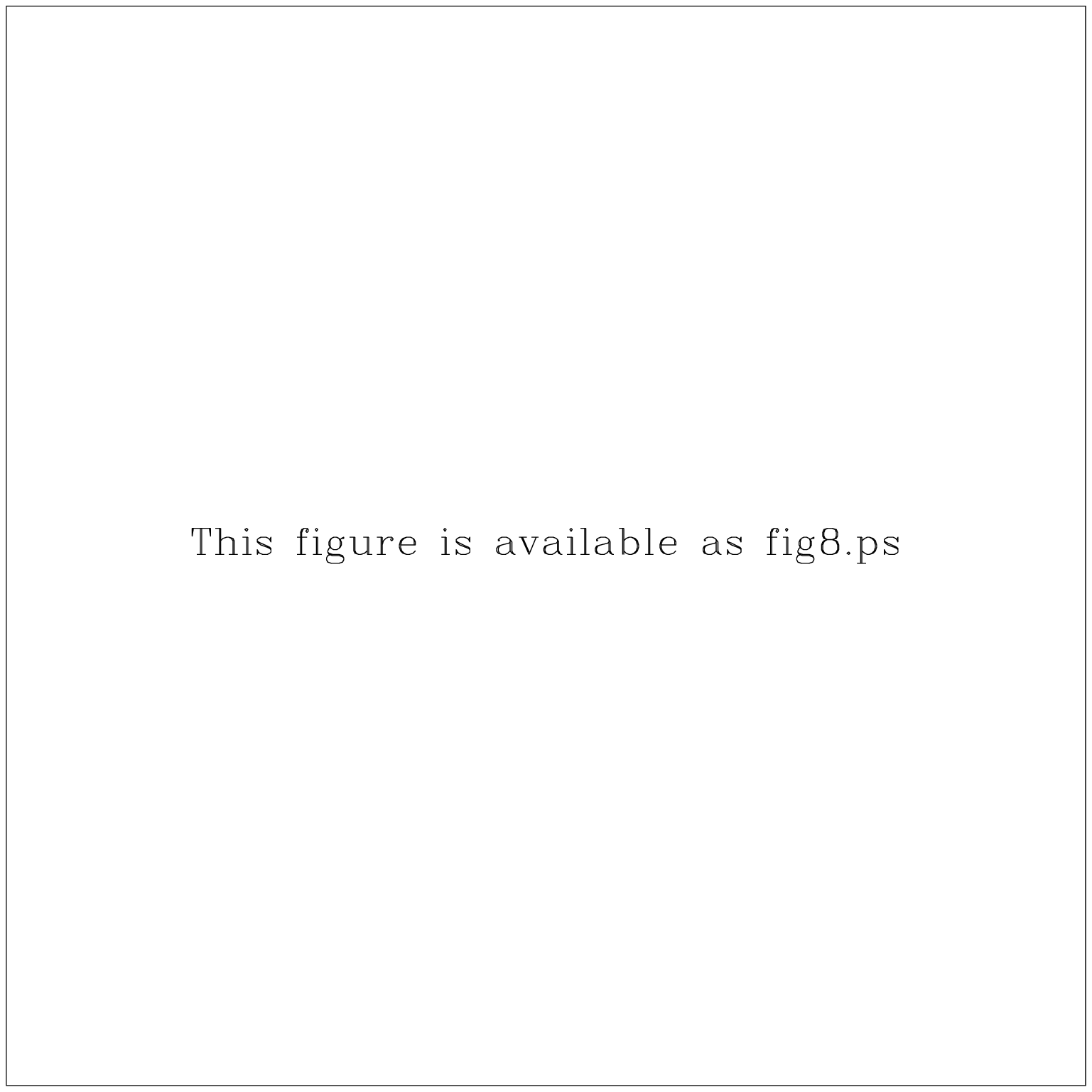}}
\caption{Plate 2. True-color image of part of  the AXAF field for which the
optical and infrared observations overlap. it is based on 6
passbands covering an extended spectral region. The blue channel is
represented by the
\u+\b\ image, the green channel by the
\v+\r\ image, and the red channel by the \j+\k\ image. This color
image covers an area of approximately $9.1\times5.1$ square arcmin
corresponding to the SUSI2-SOFI overlap in the AXAF field. }
\label{fig:plate2} 
\end{figure*}

The resulting optical and infrared mosaics in each passband are shown
in Figures~\ref{fig:opmosaic} and \ref{fig:irmosaic}.  Note that there
are no \i-band data for AXAF2. Figure~\ref{fig:plate1} is a true-color
image of the optical mosaic of the AXAF field with the blue channel
represented by the \u+\b image, the green and red channels by the \v\
and \r\ images, respectively.  Figure~\ref{fig:plate2} shows a
true-color image of the region of overlap of SOFI and SUSI2
observations with the blue channel represented by the \u+\b\ image,
the green channel by the \v+\r\ image, and the red channel by the
\j+\k\ image. All the co-added images, including the
corresponding weights and masks, are public and may be requested from
the URL "http://www.eso.org/eis".

\section {Object Catalogs}
\label{cats}

The above images were used to produce an array of source
catalogs. These include: co-added single passband, multi-passband
optical and optical-infrared catalogs of sources extracted from the
image mosaics. As in paper~I, the $\chi^2$ technique (Szalay, Connolly
\& Szolokov 1998) was used in the production of the multi-color
catalogs. These catalogs have the same format as those of paper~I and
ASCII versions can be found at the URL ``http://www.eso.org/eis''. An
overview of the available catalogs is presented in
Table~\ref{tab:summary}. Note that the number of sources in the \i\
single band catalog is smaller, because this band only includes data
from AXAF1. In the multi-color catalogs \i-band magnitudes for sources
not covered by AXAF1 are given a value of -99.9, as opposed to 99.9
for sources that were observed and detected in the $\chi^2$ image, but
not measurable in the individual band.  The coding adopted for the
multi-color catalogs in the case of no detections is described in
paper~I.

\begin{table}
\caption{Summary of Catalogs}
\begin{tabular}{lcc}
\hline\hline
Single Band  & $N_{obj}$ & Area    \\
             &           & [arcmin$^2$]\\
\hline
\u           & 3271      & 55.85   \\
\b           & 3716      & 55.85   \\
\v           & 3460      & 55.85   \\
\r           & 4137      & 55.85   \\
\i           & 1569      & 29.60   \\
\j           & 3150      & 83.26   \\
\k           & 3015      & 83.26   \\
\hline
Multi-Color   &           &         \\
\hline
AXAF Optical  & 6875      & 55.85$^1$  \\ 
AXAF OIR$^2$  & 3990      & 42.68   \\
AXAF IR       & 4583      & 83.26   \\
\hline\hline
\label{tab:summary}
\end{tabular}

\noindent
$^1$ see text \\
$^2$ OIR = optical-infrared
\end{table}

\section{Characteristics of the Data}
\label{results}

\subsection{Point-Sources}

\begin{figure}[t]
\resizebox{0.42\textwidth}{!}{\includegraphics{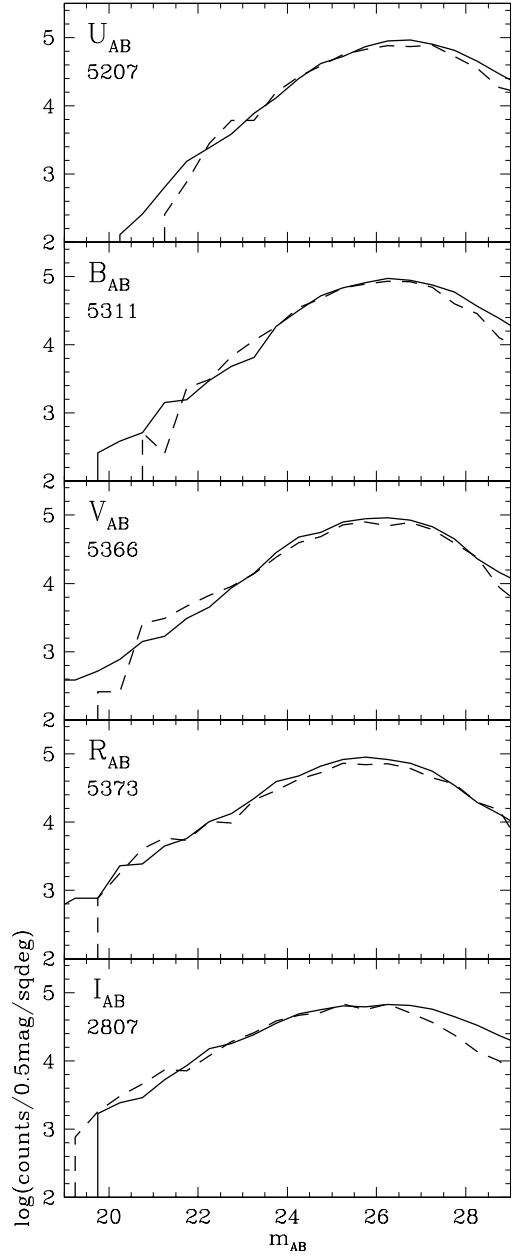}}
\caption{Optical galaxy counts derived from \u+\b+\v+\r\
$\chi^2$-image (solid line), compared with those obtained from
 EIS-DEEP HDF-S observations (dashed line). In each panel the number
 of sources measured is indicated. }
\label{fig:ncounts_comp_opt}
\end{figure}

\begin{figure}[t]
\resizebox{0.42\textwidth}{!}{\includegraphics{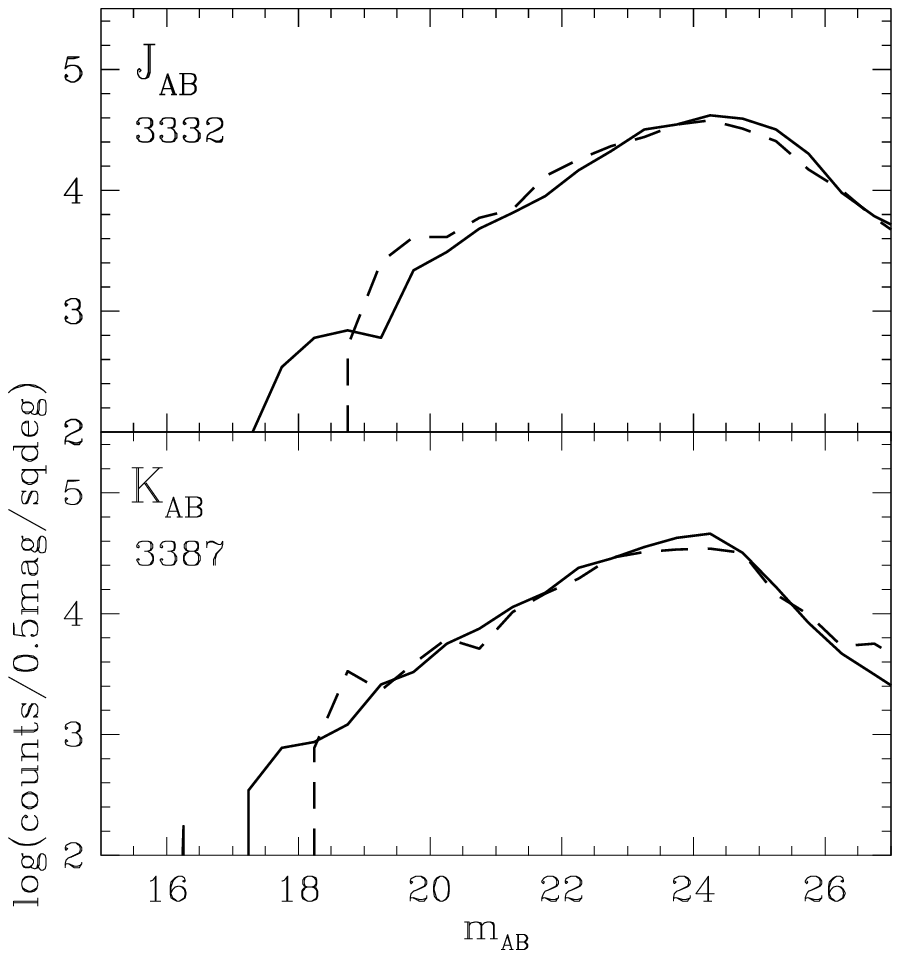}}
\caption{Infrared galaxy counts derived from the \j+\k\ $\chi^2$-image (solid
line), compared with those obtained from the EIS-DEEP HDF-S
observations (dashed line). }
\label{fig:ncounts_comp_ir}
\end{figure}

To facilitate comparisons with other data sets, throughout this
section magnitudes are expressed in the Johnson-Cousins system.
Objects with stellarity index $>0.95$ in the \v-band (chosen because
the seeing was best in this band), and $\v \lsim 22.5$, are defined to
be stars. The number of stars in the AXAF field is small, $\sim 35$,
and is in good agreement with model predictions from Haywood, Robin \&
Creze (1997, see also Paper~I).  A good check for the adopted
zeropoints is provided by the stellar color-color diagrams. Optical
color-color diagrams are presented in Figure~\ref{fig:starcol},
showing examples involving all the optical passbands used.  All the
colors have been corrected for reddening and the $(U-B)$ and $(B-V)$
colors have been corrected for the color term derived in paper~I. Only
$5\sigma$ detections, in all passbands, are included in these
diagrams.  For comparison, the empirical relations compiled by
Caldwell \etal (1993) and, whenever available, the theoretical
isochrone taken from Baraffe
\etal (1997) are used. The theoretical model assumes a
10~Gyr, [M/H]$=-1$ population, typical for halo stars, and is
fine-tuned to model low-mass main sequence stars. As can be seen, in
all cases the EIS-DEEP data are in good agreement with the empirical
and/or model sequences.

\subsection{Galaxies}

To evaluate the performance and depth of the galaxy catalogs produced
in the different optical and infrared passbands, the number counts
obtained within the AXAF region are compared with those obtained in
paper~I. For this comparison, the multi-color optical and infrared
catalogs, as detected from the $\chi^2$ image and properly normalized
to the same area, have been used. These counts are shown in
Figures~\ref{fig:ncounts_comp_opt} and \ref{fig:ncounts_comp_ir}. As
can be seen, there is a remarkable agreement between the EIS-DEEP
galaxy counts of the AXAF and the HDF-S region. The agreement of the
EIS-DEEP galaxy counts in the HDF-S region with those obtained by
other authors was already established in Paper~1.

\begin{figure*}
\resizebox{0.95\textwidth}{!}{\includegraphics{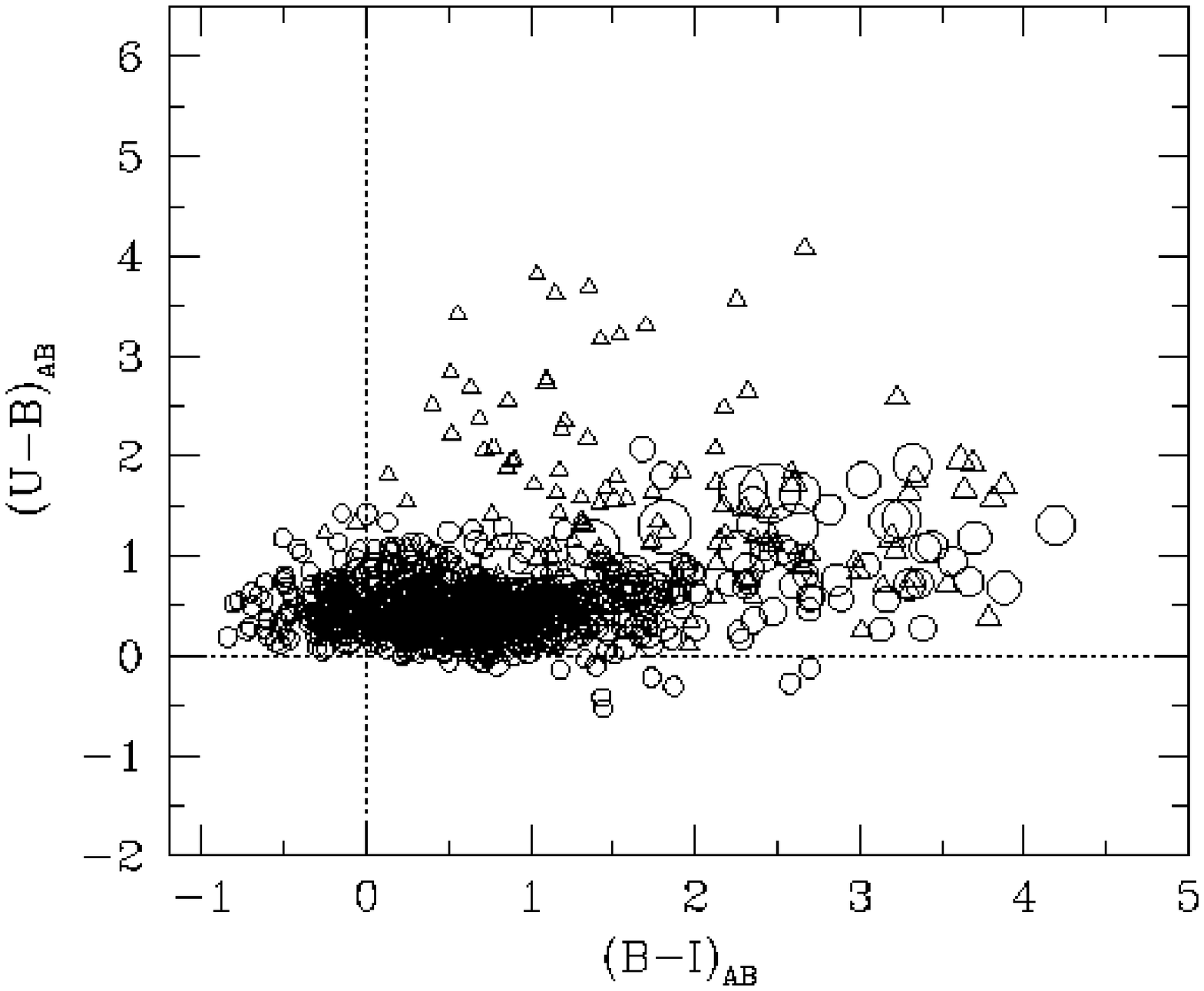}
\includegraphics{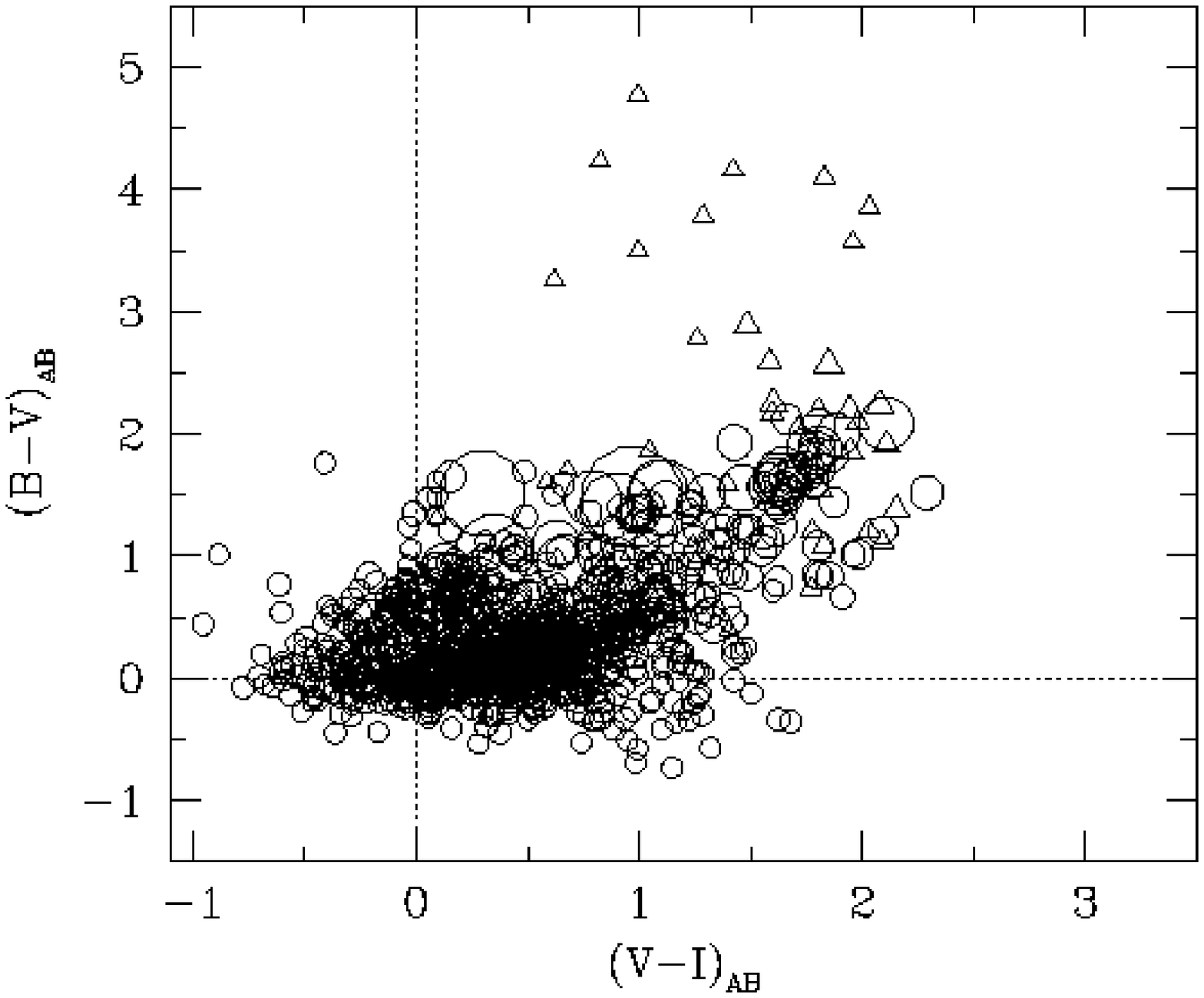}}
\caption{Galaxy $(U-B)\times(B-I)$ and $(B-V)\times(V-I)$ 
color-color diagrams. The size of
the symbols are inversely proportional to the
\i-band magnitude of the galaxy.}
\label{fig:galcol}
\end{figure*}

Since one of the primary goals of the survey has been to identify
candidate galaxies at high-z for follow-up spectroscopic observations
with the VLT, the color information available in \u, \b,
\v, and \i\ has  been used to identify preliminary Lyman-break
candidates.  Extensive work has been done to tune the color-selection
criteria for different sets of filters (\eg Steidel \& Hamilton
1993. Madau \etal 1996) and to identify regions in the color-color
diagrams populated by Lyman-break galaxies in different redshift
ranges.  Unfortunately, differences in the passbands between the SUSI2
filters and those previously used, prevent adopting the same color
criteria.  However, given the great interest in finding likely
candidates at high-z a simple approach has been adopted here for a
first cut analysis.  This was done by considering the tracks most
likely to trace the evolutionary sequence of galaxies of different
types in color-color diagrams appropriate for the SUSI2 filters
(Arnouts 1998, see also Fontana \etal 1998).  Based on these results
conservative regions in $(U-B) \times (B-I)$ and $(B-V) \times (V-I)$
diagrams, shown in Figure~\ref{fig:galcol}, were defined.  The
criteria adopted were $(U-B)_{AB}\gsim 1.5$ and $(B-I)_{AB}\lsim 2$ in
the $(U-B)
\times (B-I)$ diagram and $(B-V)_{AB}>1.5$ and $(B-V)_{AB} >
2\times(B-I)_{AB} - 0.14$. Based on the model predictions these
regions should be populated by $z>3$ galaxies.  A more precise
selection will require a more detailed analysis using the color
information to assign photometric redshifts. This will certainly be
pursued by several groups using the public data.

The galaxies shown in Figure~\ref{fig:galcol} are detections ($I_{AB}
\le 25.5$, as measured in a 3 arcsec diameter aperture) obtained
from the optical $\chi^2$-image, thus allowing for the identification
of objects that may be undetected in one or more passbands.  The
objects shown in the color-color diagrams follow several
constraints. First, they have to be $\ge 2\sigma$ detections at least
in $B$ and $I$ for the diagram shown in the left panel and $V$ and $I$
for that in the right panel.  If the object is less than a $2\sigma$
detection in the bluest passband, it is represented by a triangle,
otherwise by a circle.  For blue drop\-outs the magnitude measured by
SExtractor was used regardless of its error. If the magnitude is not
measurable (m=99.9), the 2$\sigma$ limiting magnitude is assigned to
the object.

Adopting the color selection criteria described above, galaxies in the
AXAF1 field were selected. All candidates were visually inspected and
29 \u-dropout and 13 \b-dropout galaxies remained.  These are listed
in Tables~\ref{tab:hiz_u} and \ref{tab:hiz_b}, which provide the
following information: in columns (1) and (2) right ascension and
declination (J2000.0); in column (3) the $I_{AB}$ magnitude measured
in an aperture of 3 arcsec. The remaining columns give the colors and
their respective errors. Whenever, an object is not measurable in a
given passband or the $S/N<1$ the error in the color involving this
filter is set to -1. Even though the selection criteria adopted are
admittedly crude, inspection of the selected objects indicates that
they are by and large promising.

\begin{table*}[t]
\caption{High-z Galaxy Candidates (U-dropouts).}
\label{tab:hiz_u}
\input table.ubi.tex

\end{table*}

\begin{table*}
\caption{High-z Galaxy Candidates (B-dropouts).}
\label{tab:hiz_b}
\input table.bvi.tex
\end{table*}

\section{Summary}
\label{sum}

This paper presents the results of deep optical ($\sim$ 56 square
arcmin) and infrared ($\sim$ 83 square arcmin) imaging of a region in
the southern hemisphere of very low HI column density. An area of 48
square arcmin has been observed in at least six passbands, two of
which are infrared. The observations were carried out as part of the
EIS public survey. In addition, single-passband catalogs have been
prepared as well as multi-color optical, infrared and optical-infrared
catalogs. The latter were produced using the $\chi^2$ technique, which
has proven to yield robust detections of very faint galaxies.  The
color information has been used to find possible high-z galaxies for
follow-up observations with the VLT.

The released data represent the last set of the second part of the EIS
program, involving deep imaging (EIS-DEEP). The project, as originally
envisoned, has been completed. Sixteen months after the beginning of
observations, it has produced a large volume of data from a moderately
deep, large solid angle survey (EIS-WIDE) and multi-band deep
observations of specific areas of interest. The data have been
completely reduced, source catalogs have been extracted, and
preliminary lists of potentially interesting targets have been
produced. All the data presented here and in previous papers of the
series, in the form of images, source catalogs and selected targets,
are public and can be requested at "http://www.eso.org/eis''.

\begin{acknowledgements}

We thank all the people directly or indirectly involved in the ESO
Imaging Survey effort. In particular, all the members of the EIS Working
Group (S. Charlot, G. Chincarini, S. Cristiani, J. Krautter, K. Kuijken,
K. Meisenheimer, D. Mera, Y. Mellier, M. Ramella, H. R\"ottering, R.
Saglia and P. Schneider) for the innumerable suggestions and criticisms,
the ESO OPC, the NTT team, in particular the night assistants, the ESO
Archive Groups and ECF. We would also like to thank S. Arnouts, J.
Caldwell, N. Devillard, A. Fontana, P. Rosati and R. Fosbury for their help and
assistance.  Special thanks to Riccardo Giacconi for making this effort
possible.
\end{acknowledgements}

\newpage

\newpage

\end{document}

%% file: table.ubi.tex
\begin{tabular}
{rccc
r@{\extracolsep{2mm}}r@{\hspace{1mm}}
r@{\extracolsep{2mm}}r@{\hspace{1mm}}
r@{\extracolsep{2mm}}r@{\hspace{1mm}}
r@{\extracolsep{2mm}}r@{\hspace{1mm}}}
\hline\hline\noalign{\smallskip}
\# & 
$\alpha$ &
$\delta$ &
$I_{AB}(3'')$ & 
$(U-B)$ & $\epsilon$ &
$(B-I)$ & $\epsilon$ \\
\noalign{\smallskip}\hline\noalign{\smallskip}
    1 & 03:32:05.84 & $-27$:48:15.9 &  $22.99$ &  $ 2.07$ &  $ 0.50$ &  $ 1.68$ &  $ 0.09$ \\  
    2 & 03:32:18.29 & $-27$:47:22.8 &  $24.85$ &  $ 3.63$ &  $ 5.89$ &  $ 1.15$ &  $ 0.26$ \\  
    3 & 03:32:15.15 & $-27$:47:54.5 &  $23.81$ &  $ 1.56$ &  $ 0.62$ &  $ 1.51$ &  $ 0.17$ \\  
    4 & 03:32:17.25 & $-27$:47:54.3 &  $25.26$ &  $ 1.78$ &  $ 2.41$ &  $ 1.52$ &  $ 0.46$ \\  
    5 & 03:32:09.60 & $-27$:47:39.8 &  $25.41$ &  $ 2.51$ &  $ 1.78$ &  $ 0.40$ &  $ 0.31$ \\  
    6 & 03:32:19.43 & $-27$:47:28.3 &  $25.45$ &  $ 3.82$ &  $-1.00$ &  $ 1.04$ &  $ 0.40$ \\  
    7 & 03:32:03.53 & $-27$:47:21.0 &  $24.98$ &  $ 1.68$ &  $ 2.05$ &  $ 1.45$ &  $ 0.42$ \\  
    8 & 03:32:13.53 & $-27$:47:19.6 &  $25.35$ &  $ 3.31$ &  $-1.00$ &  $ 1.70$ &  $ 0.56$ \\  
    9 & 03:32:18.30 & $-27$:47:14.5 &  $25.28$ &  $ 1.95$ &  $ 1.59$ &  $ 0.88$ &  $ 0.32$ \\  
   10 & 03:32:22.72 & $-27$:46:37.6 &  $25.36$ &  $ 3.42$ &  $ 4.47$ &  $ 0.55$ &  $ 0.30$ \\  
   11 & 03:32:15.47 & $-27$:46:31.4 &  $23.92$ &  $ 1.84$ &  $ 1.15$ &  $ 1.91$ &  $ 0.23$ \\  
   12 & 03:32:04.40 & $-27$:46:03.2 &  $25.38$ &  $ 2.69$ &  $ 2.69$ &  $ 0.63$ &  $ 0.33$ \\  
   13 & 03:32:17.93 & $-27$:45:36.1 &  $24.83$ &  $ 1.64$ &  $ 1.64$ &  $ 1.74$ &  $ 0.37$ \\  
   14 & 03:32:09.07 & $-27$:45:35.2 &  $24.99$ &  $ 2.55$ &  $ 1.90$ &  $ 0.86$ &  $ 0.26$ \\  
   15 & 03:32:16.63 & $-27$:45:20.1 &  $25.34$ &  $ 2.09$ &  $ 1.97$ &  $ 0.77$ &  $ 0.38$ \\  
   16 & 03:32:03.31 & $-27$:45:18.7 &  $25.37$ &  $ 1.81$ &  $ 1.29$ &  $ 0.13$ &  $ 0.38$ \\  
   17 & 03:32:13.79 & $-27$:45:12.9 &  $25.16$ &  $ 1.88$ &  $ 1.31$ &  $ 0.86$ &  $ 0.33$ \\  
   18 & 03:32:15.70 & $-27$:45:15.2 &  $25.38$ &  $ 1.96$ &  $ 1.97$ &  $ 0.90$ &  $ 0.43$ \\  
   19 & 03:32:09.62 & $-27$:45:14.5 &  $25.02$ &  $ 1.85$ &  $ 1.42$ &  $ 1.17$ &  $ 0.31$ \\  
   20 & 03:32:22.48 & $-27$:44:38.2 &  $25.47$ &  $ 3.21$ &  $-1.00$ &  $ 1.54$ &  $ 0.55$ \\  
   21 & 03:32:05.01 & $-27$:44:31.6 &  $24.02$ &  $ 2.17$ &  $ 1.29$ &  $ 1.34$ &  $ 0.16$ \\  
   22 & 03:32:28.26 & $-27$:44:03.4 &  $25.30$ &  $ 3.69$ &  $-1.00$ &  $ 1.35$ &  $ 0.38$ \\  
   23 & 03:32:14.49 & $-27$:44:05.6 &  $24.91$ &  $ 1.72$ &  $ 1.30$ &  $ 1.02$ &  $ 0.29$ \\  
   24 & 03:32:18.23 & $-27$:44:21.9 &  $25.18$ &  $ 1.58$ &  $ 1.68$ &  $ 1.31$ &  $ 0.38$ \\  
   25 & 03:32:03.85 & $-27$:44:05.4 &  $25.49$ &  $ 2.84$ &  $ 3.94$ &  $ 0.51$ &  $ 0.41$ \\  
   26 & 03:32:08.51 & $-27$:43:58.2 &  $25.28$ &  $ 1.54$ &  $ 0.70$ &  $ 0.25$ &  $ 0.28$ \\  
   27 & 03:32:26.39 & $-27$:43:46.5 &  $25.20$ &  $ 1.57$ &  $ 1.70$ &  $ 1.58$ &  $ 0.36$ \\  
   28 & 03:32:03.58 & $-27$:43:40.5 &  $25.03$ &  $ 2.35$ &  $ 3.77$ &  $ 1.21$ &  $ 0.43$ \\  
   29 & 03:32:16.15 & $-27$:44:01.6 &  $25.22$ &  $ 2.26$ &  $ 3.26$ &  $ 1.19$ &  $ 0.44$ \\  
 \hline\hline
\end{tabular}

%% file: table.bvi.tex
\begin{tabular}
{rccc
r@{\extracolsep{2mm}}r@{\hspace{1mm}}
r@{\extracolsep{2mm}}r@{\hspace{1mm}}
r@{\extracolsep{2mm}}r@{\hspace{1mm}}
r@{\extracolsep{2mm}}r@{\hspace{1mm}}}
\hline\hline\noalign{\smallskip}
\# & 
$\alpha$ &
$\delta$ &
$I_{AB}(3'')$ & 
$(B-V)$ & $\epsilon$ &
$(V-I)$ & $\epsilon$ \\
\noalign{\smallskip}\hline\noalign{\smallskip}

    1 & 03:32:14.37 & $-27$:48:31.2 &  $25.28$ &  $ 3.79$ &  $-1.00$ &  $ 1.29$ &  $ 0.51$ \\  
    2 & 03:32:06.62 & $-27$:47:47.5 &  $24.72$ &  $ 4.77$ &  $-1.00$ &  $ 1.00$ &  $ 0.23$ \\  
    3 & 03:32:16.21 & $-27$:47:36.6 &  $25.38$ &  $ 1.71$ &  $ 0.97$ &  $ 0.68$ &  $ 0.36$ \\  
    4 & 03:32:13.53 & $-27$:47:19.6 &  $25.35$ &  $ 1.61$ &  $ 0.53$ &  $ 0.09$ &  $ 0.31$ \\  
    5 & 03:32:20.31 & $-27$:47:18.1 &  $24.92$ &  $ 4.16$ &  $-1.00$ &  $ 1.42$ &  $ 0.34$ \\  
    6 & 03:32:09.19 & $-27$:46:52.7 &  $24.57$ &  $ 4.10$ &  $-1.00$ &  $ 1.83$ &  $ 0.37$ \\  
    7 & 03:32:05.08 & $-27$:46:12.3 &  $24.92$ &  $ 3.26$ &  $ 2.39$ &  $ 0.62$ &  $ 0.23$ \\  
    8 & 03:32:12.30 & $-27$:45:24.7 &  $23.55$ &  $ 2.89$ &  $ 1.10$ &  $ 1.48$ &  $ 0.13$ \\  
    9 & 03:32:18.94 & $-27$:45:23.6 &  $24.55$ &  $ 1.59$ &  $ 0.40$ &  $ 0.66$ &  $ 0.17$ \\  
   10 & 03:32:06.24 & $-27$:45:01.8 &  $25.07$ &  $ 1.61$ &  $ 0.65$ &  $ 0.58$ &  $ 0.27$ \\  
   11 & 03:32:25.82 & $-27$:44:34.6 &  $24.92$ &  $ 1.51$ &  $ 0.40$ &  $ 0.62$ &  $ 0.16$ \\  
   12 & 03:32:10.14 & $-27$:44:09.7 &  $24.65$ &  $ 1.69$ &  $ 0.45$ &  $ 0.49$ &  $ 0.19$ \\  
   13 & 03:32:17.75 & $-27$:43:40.1 &  $25.33$ &  $ 4.23$ &  $-1.00$ &  $ 0.83$ &  $ 0.36$ \\  

\hline\hline
\end{tabular}